\documentclass{JFM-FLM_Au}

\usepackage{color}
\newcommand{\edit}[1]{{\color{black}#1}}

\usepackage{array}
\usepackage{makecell} 

\lefttitle{D. M. Harris and others}
\righttitle{Journal of Fluid Mechanics}

\title{Bouncing to coalescence transition for droplet impact onto moving liquid pools}

\author{Daniel M. Harris$^{1}$, Luke F.L. Alventosa$^{1}$, Oliver Sand$^{1}$, Eli Silver$^{1}$, Arman Mohammadi$^{1}$, Thomas C. Sykes$^{2,3}$, Alfonso A. Castrejón-Pita$^{3}$, \and Radu Cimpeanu$^{4}$}

\affiliation{$^{1}$ School of Engineering, Brown University, Providence, RI 02912, USA \\ $^{2}$ School of Engineering, University of Warwick, Coventry CV4 7AL, UK \\ $^{3}$ Department of Engineering Science, University of Oxford, Oxford OX1 3PJ, UK \\ $^{4}$ Mathematics Institute, University of Warwick, Coventry CV4 7AL, UK}

\corresau{Daniel M. Harris, \email{daniel\_harris3@brown.edu}}

\begin{document}
\maketitle

\begin{abstract}
A droplet impacting a deep fluid bath is as common as rain over the ocean.  If the impact is sufficiently gentle, the mediating air layer remains intact, and the droplet may rebound completely from the interface.  In this work, we experimentally investigate the role of translational bath motion on the bouncing to coalescence transition.  Over a range of parameters, we find that the relative bath motion systematically decreases the normal Weber number required to transition from bouncing to merging.  Direct numerical simulations demonstrate that the depression created during impact combined with the translational motion of the bath enhances the air layer drainage on the upstream side of the droplet, ultimately favoring coalescence.  A simple geometric argument is presented that rationalizes the collapse of the experimental threshold data, extending what is known for the case of axisymmetric normal impacts to the more general 3D scenario of interest herein.

\end{abstract}

\section{Introduction}
\label{sec:Intro}
When two volumes of fluid approach one another in a gas, whether they be droplets, jets, or a droplet and a bath, the outcome of the impact is determined by the properties and persistence of a thin interstitial gas layer \citep{rein1993phenomena,neitzel2002noncoalescence,yarin2006drop,sprittles2024gas}.  For relatively low-energy collisions in air, the liquids may rebound completely from one another having never made direct contact, provided the inertia of the liquid is insufficient to drain the air layer during the interaction.  The critical role of the air layer was first hypothesized by Lord Rayleigh for the case of colliding jets, similarly framing the essential question as to ``...whether the air can be anywhere squeezed out
during the short time over which the collision extends'' \citep{rayleigh1899xxxvi}.  More specifically, the air layer must drain to a sufficient thickness such that van der Waals forces can initiate coalescence, a length scale typically on the order of 0.1 $\mu$m \citep{couder2005bouncing,tang2019bouncing,sprittles2024gas}. 

\begin{figure}
  \centerline{\includegraphics[width=135mm]{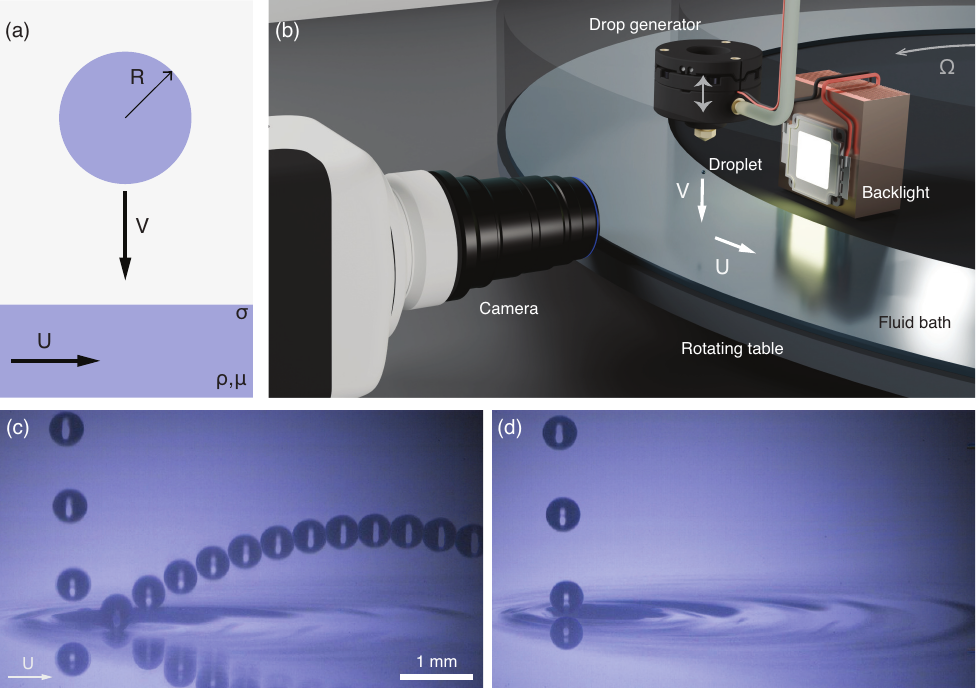}}
  \caption{(a) Schematic of a droplet impacting a moving bath of the same fluid. (b) Rendering of experimental setup.  A droplet is generated by a piezoelectric droplet generator and impacts a moving fluid layer atop a rotating table.  The dynamics are filmed from the side with a high-speed video camera.  (c,d) A 2 cSt silicone oil droplet of radius $R=0.230 \pm 0.006$ mm ($Bo=0.024$, $Oh=0.028$) impacts a fluid bath moving with horizontal speed $U=35$ cm/s. Images spaced by 1/540 seconds are directly superimposed. (c) With an impact velocity of $V=59.3 \pm 0.9$ cm/s, the droplet rebounds from the fluid bath while obtaining a horizontal velocity from the bath during contact.  (d) At $V=60.7\pm0.8$ cm/s, the droplet merges with the bath and the residual interfacial disturbance is transported downstream following coalescence.  Corresponding Videos available in Supplementary Movie 1.}
\label{fig:Fig1}
\end{figure}

In the present work, we focus on the bouncing to coalescence (BC) transition for the case of a droplet impacting a deep planar bath, first examined in the seminal contribution of \cite{schotland1960experimental}. Schotland identified the transition between rebound and coalescence for a continuous stream of droplets impacting a fluid bath at an oblique angle.  They noted that for a fixed liquid and gas pair, the transition between rebound and coalescence was principally dictated by the normal Weber number, defined as

\begin{equation}
We=\frac{\rho V^2 R}{\sigma},
\end{equation}
where $\rho$ is the fluid density, $V$ is the droplet impact velocity projected normal to the undisturbed bath interface, $R$ is the droplet radius, and $\sigma$ is the surface tension (Figure \ref{fig:Fig1}(a)).  The bouncing to coalescence transition was further interrogated for oblique streams of droplets in numerous follow-up works \citep{jayaratne1964coalescence,ching1984droplet,zhbankova1990collision,doak2016rebound}, where various hypotheses about the possible role of relative tangential motion have been proposed, but have hitherto lacked any experimental backing.

Single droplets may also rebound from a quiescent fluid bath, with the bouncing to coalescence transition occurring above a critical Weber number that depends on the liquid and gas properties \citep{rodriguez1985some,pan2007dynamics,huang2008study,zhao2011transition,zou2011experimental,tang2016nonmonotonic,tang2018bouncing,wu2020small,wu2022droplet}.  Although the depth ($H$) of the fluid bath also plays a critical role in determining the outcome when it is on the order of the droplet size \citep{pan2007dynamics}, our focus here is on the case of deep baths where $H \gg R$.  Despite the considerable progress over the last several decades, these prior works have focused exclusively on the axisymmetric case of normal droplet impacts on a still fluid bath.  At this point it remains unknown what role tangential motion might play in determining the transition between bouncing and coalescence, and in particular, if the normal Weber number is sufficient to determine the outcome.  Herein we experimentally demonstrate that relative bath motion leads to a systematic reduction in the critical normal Weber number. This finding is in contrast to other scenarios where relatively rapid substrate translation has been shown to favor non-coalesce behavior \citep{neitzel2002noncoalescence,che2015impact,castrejon2016droplet,gauthier2016aerodynamic,sprittles2024gas}, although in some scenarios a small radial boundary flow has been shown to enhance air-layer drainage and accelerate coalescence \citep{lo2017mechanism}.  Companion three-dimensional direct numerical simulation results allow us to interrogate the air layer dynamics during impact and isolate the mechanism responsible for the reduction.

\section{Experimental methods}
\label{sec:Experiment}

\subsection{Experimental setup}
A rendering of the experimental setup is shown in Figure \ref{fig:Fig1}(b).  Droplets of silicone oil are generated from a piezoelectric droplet-on-demand generator \citep{harris2015low,ionkin2018note} and impact a fluid bath of the same fluid.  The working fluids are silicone oil (Clearco Products) with kinematic viscosities $\nu = \mu/\rho=$ 2, 20, and 50 cSt.  These fluids correspond (respectively) to densities of $\rho=0.873$, 0.949, 0.960 g/cm$^3$ and surface tension coefficients of $\sigma=18.7$, 20.6, and 20.8 dynes/cm, as provided by the manufacturer datasheet.  The latest version of the generator, and the one employed for this work, is documented at \url{https://github.com/harrislab-brown/DropGen}.  Interchangeable nozzles allow for control of the droplet radius \-- commercial 3D-printer nozzles of outlet diameter $0.3, 0.5$ and $1.0$ mm are used in the present work, resulting in droplet sizes ranging from $R=0.17-0.44$ mm. Smaller droplets ($R\approx0.1$ mm) were successfully produced with a $0.2$ mm diameter nozzle, however the BC transition velocity could not be exceeded with gravity alone, and were only ever observed to bounce at their limiting terminal velocity. Apart from very small droplets ($R \lesssim 100$ $\mu$m), the BC transition has been documented to be largely insensitive to the droplet radius  \citep{schotland1960experimental,pan2007dynamics,lewin2024liquid}. The fluid and droplet sizes used correspond to Bond numbers ($Bo=\rho g R^2/\sigma$) ranging from 0.01 to 0.09 and Ohnesorge numbers ($Oh=\mu/\sqrt{\rho \sigma R}$) ranging from 0.02 to 0.5, suggesting that gravity and viscosity play a secondary role with the dynamics dominated by inertia and capillarity \citep{alventosa2023inertio}.  The droplet generator head is mounted on a vertically oriented motorized translation stage controlled by a stepper motor that allows for vertical displacement of the generator (with resolution of 1280 steps per mm) in order to vary the droplet impact velocity, and therein the Weber number.  

Horizontal motion of the liquid bath is achieved via a large rotating table, with design and construction inspired by the `Weather in a Tank' apparatus \citep{illari2009weather}. The rotating circular platform has a diameter of $0.71$ m and is driven into steady rotation at speeds up to $\Omega=10$ rpm by a friction wheel. The fluid is confined within an annulus with inner diameter $0.61$ m surrounded on either side by thin clear walls of height $7.6$ cm and thickness $4.7$ mm. A similar rotating annular bath setup has been used in prior work to study the rebound of vertical fluid jets impacting a translating fluid bath \citep{thrasher2007bouncing}.  The annulus is filled with 1 liter of silicone oil corresponding to a mean fluid depth of 4.5 mm. Prior work has demonstrated that depths greater than approximately 2-3 droplet radii correspond to deep-bath limit for the bouncing to coalescence threshold \citep{pan2007dynamics,tang2018bouncing,wu2022droplet}, the regime of interest for the present work. The droplet generator is positioned such that each droplet impacts at the center of the annular region.  Bath linear translation speeds (as evaluated at the impact location) of up to $U=35$ cm/s are achieved with the design and explored in the present work. The speed is monitored by an encoder attached to a second spring-loaded friction wheel and held constant via closed-loop feedback control. Maximum uncertainty in the bath speed $U$ at the impact position was estimated to be $2\%$ of the reported value. Inevitably, the table rotation leads to a curved parabolic interface shape in steady state.  At the maximum speed used in the present work, the interface has an estimated $2.2^{\circ}$ inward slope at the impact location.  The annular bath is covered when not in use to minimize ambient contamination. We note that in the reference frame of the bath, the problem is one of an oblique droplet impact with an incident angle relative to the horizontal of $\tan\theta=V/U$, albeit with potentially different air boundary layer dynamics.

The dynamics are imaged through the outer annulus with a high-speed video camera (Phantom Miro LC 311) and macro lens (Laowa 25 mm Ultra Macro) at a frame rate of 15,000 frames per second, exposure time of $50$ $\mu$s, and spatial resolution of 7.8 $\mu$m per pixel.  The droplet and bath are illuminated using a custom LED backlight.  To clearly view the droplet and bath interface in all cases the camera is angled downward at $8^{\circ}$, leading to small errors on distances and speeds of no more than $1\%$.  A transition from bouncing to coalescence following a slight increase in $V$ is shown in Figure \ref{fig:Fig1}(c,d).

\subsection{Experimental procedure}

To start an experiment, the bath is set at the desired linear speed and left to equilibrate for several minutes before experiments are performed.  Droplets are then generated at a frequency of 2 Hz or slower.  The generator is positioned sufficiently high  such that all droplets coalesce.  The generator is then lowered by 1.0 mm and 15 successive droplet impacts are observed.  The lowering process is repeated until at least 13 of 15 of the observed droplets rebound, at which point a single video recording is taken for droplet size and velocity measurements.  The generator is then raised by 1.0 mm until 13 of 15 droplets coalesce, and another video is taken.  This process (incremental lowering/raising) is repeated twice more before moving to a new table speed.  After surveying all table speeds, the overall procedure is repeated twice more for a total of three times.  This protocol results in a total of nine measurements of the bouncing threshold and nine measurements of the coalescence threshold at each table speed. The criterion of 13 of 15 was chosen to robustly deal with outliers that occur even away from the transition point, presumably resulting from ambient vibrations, air currents, surface contamination, and other inevitable imperfections. The transition between regimes occurs over a relatively narrow range of impact velocities, typically on the order of 2 cm/s.  The video recordings are post-processed in MATLAB using Canny edge detection to obtain the droplet radius and velocities as in prior work \citep{galeano2021capillary,alventosa2023inertio}. In the experiment, the contact time $t_c$ is defined as the duration between the frame where the highest point of the droplet is one diameter above the undisturbed free surface (i.e. $z=2R$) to the frame where it returns to that same height. For computation of the vertical impact velocity, a parabola is fit to at least 25 points prior to impact for the vertical location of the highest point of the droplet, $z(t)$.  The slope of the fitted parabola is then evaluated at the impact frame (defined as $t=0$), yielding the incident vertical (normal) velocity $V$.  A similar protocol is followed for the horizontal velocity, where a line is fit to the horizontal location of the highest point of the droplet $x(t)$, and derivative evaluated at $t=0$ yielding the incident horizontal velocity $U_i$. On average, $U_i$ was less than 9\% of the pool speed $U$, and likely a consequence of the air boundary layer driving some lateral motion prior to impact. The same procedure was followed for the outgoing vertical ($V_e$) and horizontal ($U_e$) velocities, fitting a parabola and line to $z(t)$ and $x(t)$, respectively, for at least 70 points after impact, and evaluating the derivatives at $t=t_c$. The vertical coefficient of restitution was then computed as $\alpha=V_e/V$ and the horizontal coefficient of restitution as $\epsilon=\Delta U/U=(U_e-U_i)/U$.

\section{Direct numerical simulation}
\label{sec:DNS}

\begin{figure}
  \centerline{\includegraphics[width=135mm]{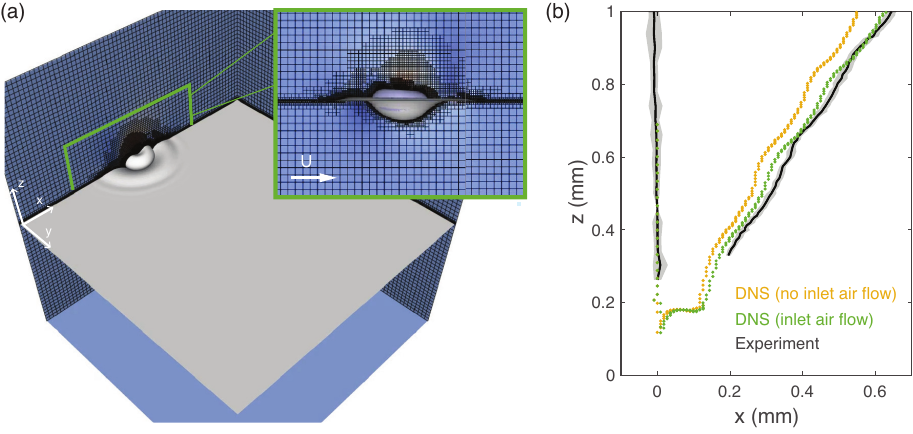}}
  \caption{(a) Three-dimensional computational domain, with adaptive mesh highlighted in the inset. Supplementary videos of representative test cases (impact onto both static and moving pools) are also made available as Supplementary Movies 2 and 3. (b) Trajectory of highest point of droplet in the symmetry plane for a case of a $2$ cSt silicone oil droplet with $R=0.23$ mm ($Bo=0.024$, $Oh=0.028$), $V=60.3$ cm/s, and $U=15$ cm/s. Yellow markers are predictions without inlet airflow, green markers are predictions with uniform inlet airflow, and solid lines are experimental measurements.  The shaded region represents two standard deviations across experimental trials.}
\label{fig:Fig2}
\end{figure}

Complementing the experimental setup deployed in this study, we have developed a three-dimensional numerical counterpart using the open-source software Basilisk \citep{popinet2009accurate, popinet2015quadtree}, which has been successfully used for studying bouncing phenomena recently \citep{alventosa2023inertio, sanjay2023does, ray2024new,phillips2025modelling}. Our multi-scale setup, balancing the requirement to resolve the thin gas film entrapped between the impacting drop and the moving pool, with a sufficiently large scale domain to ensure the rebound is unaffected by finite size effects, make it an ideal candidate for the adaptive mesh refinement and parallelization features of the code. The full implementation is made available at \url{https://github.com/rcsc-group/BouncingDropletsMovingPool3D}. The domain construction builds on a computational box measuring $20^3$ dimensionless units (relative to a dimensionless impacting drop radius of $1$), making use of the one viable symmetry plane (the $x-z$ plane), and is illustrated in Figure~\ref{fig:Fig2}(a). This choice ensures that outward propagating waves have sufficient space to develop and exit the domain cleanly. The droplet is initialized with its south pole one radius above the pool, and a uniform imposed vertical velocity field within. Gravitational effects are also included in the simulation setup. Inflow boundary conditions (and in particular a non-zero horizontal velocity in $x$) are employed across one face of the cube, with free-slip and impermeability prescribed at the bottom of the geometry, and outflow conditions on all other boundaries. The noted uniform directional horizontal velocity field is prescribed in both pool and gas regions, replicating a fully developed flow near the rotating table liquid-gas surface. The air boundary layer above the rotating fluid surface in the experiment can be estimated using the expression $\delta\approx2.5\left(\mu_a L /\left(\rho_a U\right)\right)^{1 / 2}$, where $L$ is the radial position of impact on the table, and $\mu_a$ and $\rho_a$ are the viscosity and density of air, respectively \citep{gauthier2016aerodynamic}. Across all speeds, $\delta\gtrsim 6$ cm, which is substantially larger than the computational domain and justifies the use of a uniform air inlet condition. We find that the horizontal velocity induced by the imposed non-zero horizontal air flow during the transient stage prior to impact imparts a change of less than $1\%$ in the horizontal velocity of the drop (measured at its centre of mass) relative to the initially imposed vertical velocity. The extremal case in our study, wherein the ratio between the horizontal velocity of the pool and vertical velocity of the drop is approximately $0.5$, translates to an increase of $0.371\%$ in the horizontal velocity component of the drop just prior to impact. Ensuring that the gas region has a non-zero velocity was found nevertheless to be crucial for good agreement with experiments in scenarios in which the horizontal velocity of the pool was non-negligible ($\gtrapprox 10\%$) relative to the vertical drop velocity, as showcased in Figure~\ref{fig:Fig2}(b). Furthermore, the uniform inflow condition makes our results directly applicable to the case of an oblique droplet impact, as the problem is identical when shifted into the inertial reference frame of the bath.

The setup described above, equipped with a mesh refinement strategy based on the position of the liquid-gas interfaces, as well as changes in magnitude of velocity components and vorticity, leads to tractable configurations with $O(10^6)$ computational grid cells, which are executed over $20$ dimensionless time units (typically equivalent to $O(10^{-2})$ s in dimensional terms).
This can be achieved in approximately $2000-3000$ CPU hours for each run, executed on local high performance computing clusters over $8$-$16$ CPUs.
Following careful verification and previous investigations \citep{alventosa2023inertio}, we found that a minimum cell size measuring $O(1)\ \mu$m (typically $4.49\ \mu$m when using $2^{10}$ grid cells per dimension as the most refined cell size) is sufficient to ensure mesh-independent results in the target metrics. Appendix~\ref{app:DNS_Verification} includes a detailed mesh resolution study that underpins this conclusion. Classical quantities of interest such as contact time and coefficients of restitution are shown to converge at this level of refinement, while the behaviour of the gas film thickness is sufficiently robust so as to enable reliable interrogation and to support mechanistic explanations. This resolution is also well aligned with previous entrapped gas film measurements in this regime \citep{tang2019bouncing}, albeit without three-dimensional flow features. Few comprehensive numerical studies have been considered in full three-dimensional setups given the high numerical cost, however there are excellent examples in both bouncing \citep{ramirez2020lifting} and splashing \citep{wang2023analysis} contexts, with the authors highlighting the considerable resources required to fully resolve the features of interest in the target regimes across a sufficiently wide section of the parameter space.

In our implementation coalescence between the drop and the pool is prevented at the level of the volume-of-fluid method by construction (similar to the approach in \cite{ramirez2020lifting}) to avoid artificial merging that could disrupt the bouncing dynamics. With careful mesh adaptivity, this approach yields mesh-independent results but cannot naturally capture coalescence, which would require sub-continuum physics considerations \citep{lewin2024collision}. Instead, we use our numerical infrastructure to probe bouncing regimes identified experimentally, providing mechanistic insight and enabling detailed analysis of rebound dynamics and air layer evolution in this delicate regime.

\begin{figure}
  \centerline{\includegraphics[width=135mm]{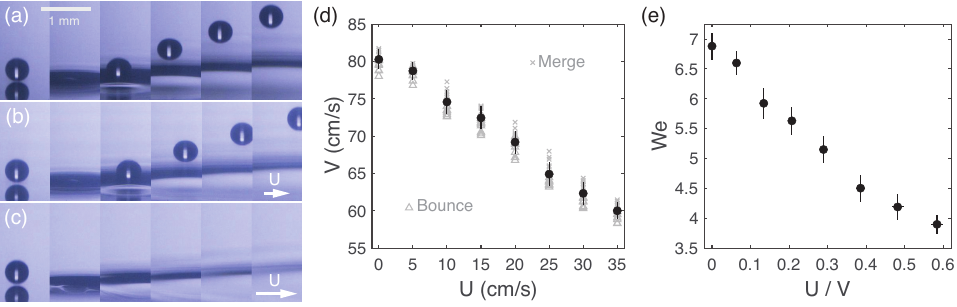}}
  \caption{Droplets (2 cSt silicone oil) of radius $R=0.230\pm 0.006$ mm \edit{($Bo=0.024$, $Oh=0.028$)} with incident vertical speed $V$ impact a fluid bath moving with horizontal speed $U$. (a-c) 
  Fixed impact velocity $V=73.1 \pm 1.0$ cm/s with increasing bath speed. Successive images in each sequence are spaced 1/750 seconds apart. Bouncing is observed for (a) $U=0$ cm/s and (b) $U=10$ cm/s, with merging at (c) $U=20$ cm/s. Corresponding videos available in Supplementary Movie 4. (d) Bouncing to coalescence transition as a function of bath speed. Gray triangles are trials where bouncing was observed, gray $\times$s are trials where coalescence was observed, and black markers are the mean transition values at each bath speed. (e) Critical $We$ as a function of normalized bath speed $U/V$. In all cases error bars represent propagated error, including one standard deviation across trials.}
\label{fig:Fig3}
\end{figure}

\section{Results}
\label{sec:Results}
Consistent with what is known from prior work on axisymmetric impacts (and showcased in Figure \ref{fig:Fig1}(c,d)), for a given set of parameters, there is a critical vertical impact velocity $V$ at which the behavior transitions from bouncing to merging.  The goal of the present work is to explore how the introduction of translational bath speed $U$ influences this critical velocity.  An experimental sequence is shown in Figure \ref{fig:Fig3}, where now the vertical velocity is fixed, and the bath speed is increased incrementally.  At $U=0$ cm/s, the droplet rebounds in an axisymmetric manner, as expected.  For $U=10$ cm/s, the droplet also successfully rebounds, but departs the surface with a non-zero horizontal velocity bestowed onto it by the bath.  However, as the speed is increased further to $U=20$ cm/s, merging ensues, suggesting that the bath motion has introduced new conditions favorable to air layer drainage.  For fixed droplet size and fluid parameters, this effect is explored systematically in Figure \ref{fig:Fig3}(d).  Individual trials for the bouncing and merging threshold are shown as the gray markers, with the combined data summarized by the black markers.  As can clearly be seen, the bath motion $U$ leads to a systematic reduction in the necessary vertical velocity required to initiate merging.  The non-dimensionalized version is presented in Figure \ref{fig:Fig3}(e), again highlighting this monotonic trend.

\begin{figure}
  \centerline{\includegraphics[width=135mm]{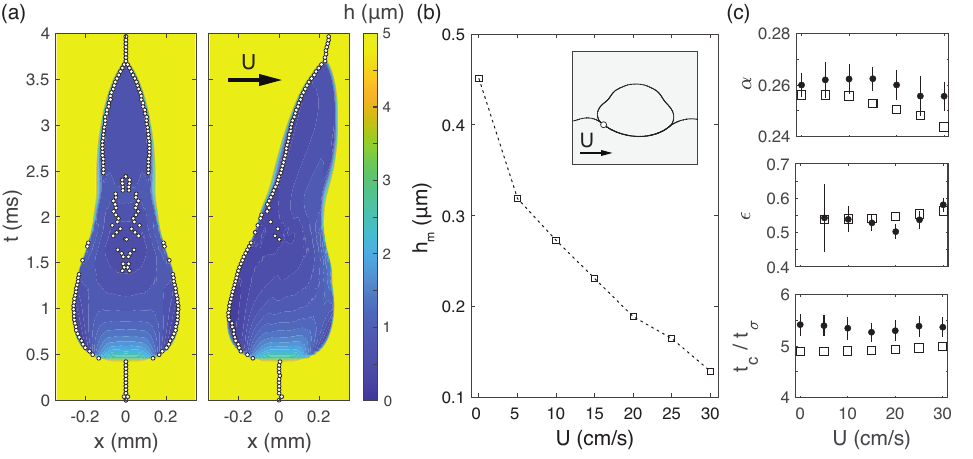}}
  \caption{Simulation results of droplets (2 cSt silicone oil) of radius $R=0.23$ mm \edit{($Bo=0.024$, $Oh=0.028$)} with incident vertical speed $V=60.3$ cm/s impacting a fluid bath moving with horizontal speed $U$. (a) Evolution of air layer thickness profile along symmetry plane for the case of $U=0$ (left) and $U=15$ cm/s (right). Circular markers indicate the point of minimum thickness, which systematically occurs on the upstream side of the contact region for $U>0$.  (b) Minimum air layer thickness as a function of bath speed $U$.  Inset shows a typical slice from the simulation, with the circular marker indicating the position of minimum thickness. (c) Vertical coefficient of restitution ($\alpha$), horizontal coefficient of restitution ($\epsilon$), and non-dimensional contact time ($t_c/t_\sigma$) as a function of $U$ for simulations ($\square$) and experiments ($\bullet$). $t_\sigma$ is the inertio-capillary timescale defined as $\sqrt{\rho R^3/\sigma}$. For experimental data, error bars represent propagated error, including one standard deviation across trials.}
\label{fig:Fig4}
\end{figure}

As reviewed in the introduction, the persistence of the intervening air layer determines the impact outcome.  As such, our experimental observations suggest that the evacuation of the air layer is enhanced by the motion of the bath.  To elucidate the physical mechanism underlying this conclusion, we turn to our companion numerical simulations.  In Figure \ref{fig:Fig4}(a), the thickness of the air layer in the symmetry plane is plotted as it evolves in time for a representative bouncing case with and without bath motion.  For the case of $U=0$, the air layer evolves symmetrically, as expected, with the minimum air gap thickness (denoted by the circular markers) occurring on the rim of the contact region for most of the contact period. The draining of the air gap exhibits rich dynamics even in this base scenario, with the draining of the gas film and the transition from ascent to descent stages leading to jumps in the location of the minimum from the rim towards the south pole of the drop and back, as illustrated during a timescale approximately described by $1.5 < t < 2.5$ ms in the leftmost panel. This delicate interplay has been reported previously in the normal impact scenario by \cite{tang2019bouncing}, withstanding significant parametric variation. Introducing bath motion breaks this symmetry, and in particular, the location of minimum thickness now systematically occurs on the upstream edge of the contact region, with a much shorter-lived switching period around $t \approx 1.8$ ms in which the minimum is found closer to the south pole of the drop. Figure \ref{fig:Fig4}(b) shows the absolute minimum thickness as a function of bath speed, demonstrating that the introduction of bath speed also decreases the minimum thickness, consistent with our conclusion that the horizontal motion must enhance air layer drainage locally, favoring merging behavior. We note that the reported minimum air gap thickness occurs at a sub-grid cell lengthscale, thereby requiring a robust post-processing algorithm to ensure that the extracted value is free from numerical interfacial reconstruction errors, is maintained for a sufficiently long timescale to be reliable, and is standardized across different test cases. An imposed criterion of the air layer thickness being extracted using a rolling window characterized by a lengthscale of at least $5$ adjacent contact grid cells and sustained over $10\%$ of the contact time was found to provide a suitable balance between avoiding numerically-induced artifacts in both space and time, and capturing the defining features of the different stages of the rich bouncing dynamics. Figure~\ref{fig:FigApp}(b) in Appendix~\ref{app:DNS_Verification} provides an illustration of the gas film dynamics focusing on the contact timescale. We argue that the measurements showcase consistent behaviour and a strong level of robustness across different mesh resolution levels while acknowledging the well-known mesh-dependent properties of the calculation, which makes quantitative predictions unreliable. We therefore aim to use these results in order to understand impact dynamics and key trends in the obtained measurements. With these considerations and while only drawing from the qualitative trends of the data therein, as depicted in the inset of Figure \ref{fig:Fig4}(b), we find that the depression created by the droplet during impact is pressed against the tilted upstream side of the droplet as the bath continues to translate, and is the mechanism responsible for the enhanced drainage.  Although not a primary objective of the article, we also present typical bouncing metrics (coefficients of restitution and contact time) for these parameters in Figure \ref{fig:Fig4}(c), demonstrating that these factors are largely unaffected by the bath motion. \citet{zhbankova1990collision} suggested, but did not demonstrate, that relative tangential velocity might prolong the contact time and ultimately promote air layer drainage. Here we show that the contact time is essentially constant with $U$, and therefore a different mechanism is at play behind the favored coalescence in our experiments.

\begin{figure}
  \centerline{\includegraphics[width=135mm]{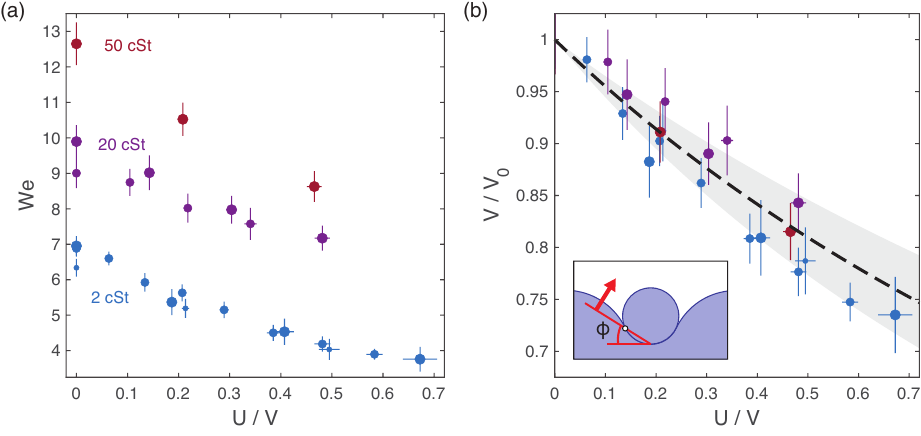}}
  \caption{(a) Bouncing to coalescence transition for silicone oil droplets of different viscosities (marker color) and radii (marker size) as a function of the normalized bath speed. For 2 cSt oil (blue): small, medium, and large markers correspond to $R=0.169 \pm 0.008$, $0.230 \pm 0.006$, and, $0.403 \pm 0.008$ mm, respectively \edit{($Bo=0.013,0.024,0.074$ and $Oh=0.033,0.028,0.022$)}.  For 20 cSt oil (purple): medium and large markers correspond to $R=0.208 \pm 0.009$ and $0.398 \pm 0.008$ mm, respectively \edit{($Bo=0.020,0.071$ and $Oh=0.30,0.22$)}. For 50 cSt oil (red): large markers correspond to $R=0.443 \pm 0.016$ mm \edit{($Bo=0.089$ and $Oh=0.51$)}. \edit{In all cases, droplets impact onto a bath of identical fluid.} (b) Critical vertical velocity $V$ normalized by the critical velocity for a still bath $V_0$ (i.e. with $U=0$) under otherwise equivalent conditions.  Dashed line (and shaded region) shows equation \ref{eqn:collapse} with film angle parameter $\phi=25.2\pm5.2^{\circ}$. \edit{The ratio $V/V_0$ can also be expressed as $\left(We/We_0\right)^{1/2}$, where $We_0$ is the critical normal Weber number for the equivalent axisymmetric case.} In all cases error bars represent propagated error, including one standard deviation across trials.}
\label{fig:Fig5}
\end{figure}

Lastly, we explore the generality of our main finding: that the normal Weber number for the bouncing to coalescence transition decreases with relative bath motion. Figure \ref{fig:Fig5}(a) presents the results for the critical impact Weber number as a function of normalized bath speed for multiple droplet sizes and viscosities (up to 50 cSt), \edit{with the droplet and bath always being of the same fluid}.  \edit{Consistent with prior understanding in the axisymmetric case \citep{schotland1960experimental,pan2007dynamics}, the threshold is only weakly sensitive to the droplet size in this regime. We also observe a clear increase in the threshold with increased viscosity, similarly documented and rationalized in prior work on the axisymmetric problem \citep{tang2018bouncing}.}  Nevertheless, in every parameter combination explored herein, the critical $We$ systematically decreases with bath speed.  If we replot the data for the critical vertical velocity $V$, normalized by the axisymmetric critical velocity $V_0$ (for $U=0$) in each case, all of the collected data nearly collapses along a single curve in Figure \ref{fig:Fig5}(b).  \edit{Note that the ratio $V/V_0$ can also be expressed as $\left(We/We_0\right)^{1/2}$, where $We_0$ is the critical normal Weber number for the equivalent axisymmetric case.}  This collapse, which involves only kinematic parameters of the problem, suggests a geometric interpretation of our result.  As discerned from the simulation, the tilted upstream contact surface and the relative bath motion conspire to promote upstream air layer drainage.  If one then assumes air layer evacuation is caused by the normal projection of both $V$ and $U$ onto a tilted surface with angle $\phi$ (see inset of Figure \ref{fig:Fig5}(b)), a new characteristic velocity for the transition can be defined as $V_c=V\cos\phi+U\sin\phi$.  When $U=0$, $V_c=V_0\cos\phi$, and we can find

\begin{equation}
\frac{V}{V_0}=\frac{1}{1+\frac{U}{V}\tan\phi}. \label{eqn:collapse}
\end{equation}
Fitting the best value of $\phi$ to each dataset yields $\phi=25.2\pm5.2^{\circ}$, which is consistent with the typical average film slopes observed in our simulations and those of \citet{lewin2024liquid} for the axisymmetric scenario.  While the details of the gas film evolution are dynamic and complex, our combined results nevertheless suggest a simple coherent physical picture.  The deformation of the bath caused by the droplet impact itself leads to a tilted mediating air film, with drainage on the upstream side then enhanced by the relative bath motion.  Crudely approximating the substrate as a tilted planar surface with angle $\phi\approx25^{\circ}$ allows us to rationalize the collapse of all our data, motivated by this simple geometric understanding of the underlying physical mechanism at play. The successful collapse of the transition data with this geometric viewpoint suggests a revised definition of the critical Weber, $We_c=\rho (V\cos\phi+U\sin\phi)^2R/\sigma$, that generalizes beyond the axisymmetric scenario.

\section{Discussion}
\label{sec:Discussion}

Despite the progress presented herein, there are some limitations that should be highlighted and may also serve as motivation for future work.  Due to experimental constraints, the study was restricted to scenarios where $U<V$ (corresponding to near-normal impacts in the equivalent oblique problem).  It is plausible that higher bath speeds (or shallow impact angles as in \cite{leneweit2005regimes}) would result in different phenomenology.  \edit{In particular, aerodynamic levitation of droplets has been documented in cases with higher substrate speeds ($U\gtrsim 1$ m/s) \citep{gauthier2016aerodynamic,castrejon2016droplet}, which suggests the possibility of a non-monotonic trend in the critical Weber number.} Furthermore, the present work focuses on millimetric droplets, whereas it has been shown that for the case of both oblique droplet streams and single droplets impacting normally, small droplets ($R\lesssim 100$ $\mu$m) exhibit an additional coalescence regime that precedes the bouncing regime \citep{jayaratne1964coalescence,bach2004coalescence,zhao2011transition,lewin2024liquid}, leading to a coalescence-bouncing-coalescence transition sequence as the Weber number is increased from zero. Future work on the topic might also explore the role of layer depth \citep{pan2007dynamics}, charge \citep{jayaratne1964coalescence}, differing drop and bath fluids \citep{wu2022droplet}, ambient gas properties \citep{schotland1960experimental}, wind \citep{liu2018experimental}, curved interfaces (such as sessile droplets \citep{moon2018observation}), or unsteady fluid interfaces \citep{couder2005bouncing,che2015impact}.  \edit{In particular, future experiments with differing bath and droplet viscosities would be particularly valuable towards further elucidating the subtle role of substrate deformation in defining the BC transition.}  From the modeling perspective, additional work extending the hybrid open-source computational framework of \cite{sprittles2024gas} and \cite{lewin2024liquid} to 3D would provide additional physical insight and a predictive capability currently unavailable for this highly multiscale problem.

\appendix

\section{Numerical resolution study}\label{app:DNS_Verification}

The numerical study in this investigation represents a non-trivial and computationally resource-intensive campaign that greatly benefits from the capabilities of the underlying algorithmic setup in Basilisk.
In Section~\ref{sec:DNS} we have introduced key structural elements of our implementation. In what follows we expand on performance and result robustness aspects, with the reference test case previously illustrated in Figure~\ref{fig:Fig2} and the surrounding discussions as its basis. This case represents an impact scenario onto a pool moving at a moderate speed (halfway through the parameter sweep in our study for the variable in question), readily leading to the (symmetry-breaking) behaviour of interest in the three-dimensional flow.

\begin{figure}
  \includegraphics[width=135mm]{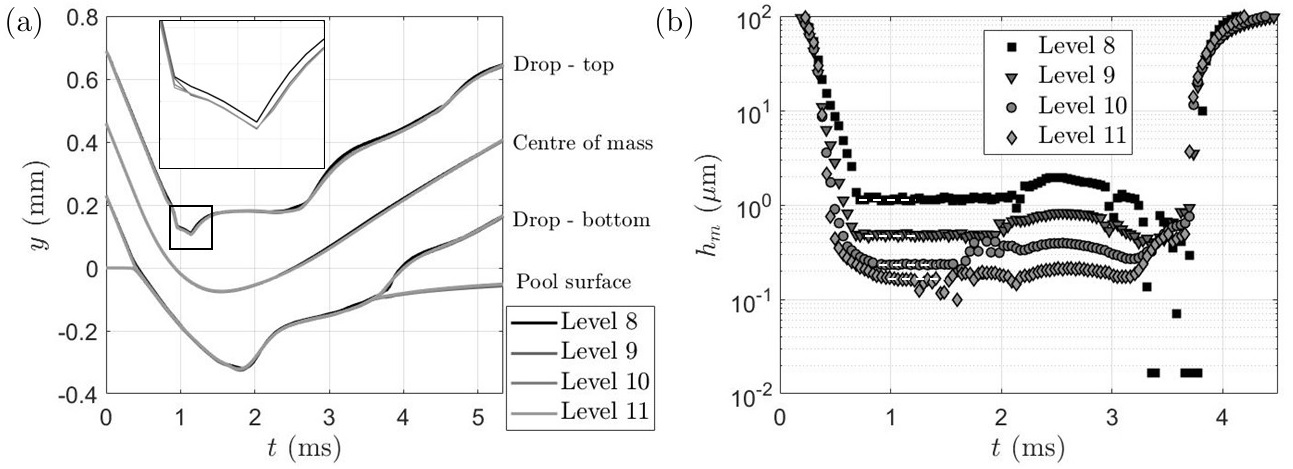}
  \caption{Resolution study for the case of a $2$ cSt silicone oil droplet with $R=0.23$ mm, $V=60.3$ cm/s, and $U=15$ cm/s, as previously presented in Figure~\ref{fig:Fig2}. Four resolution levels, with minimum grid cell sizes ranging from $17.969\ \mu$m (Level 8) down to $2.246\ \mu$m (Level 11), and further properties detailed in Table~\ref{tab:tabApp}, are considered. (a) Interfacial dynamics of drop vertical coordinate $y$ extracted at the plane of symmetry with the droplet minimum and maximum coordinates, as well as its center of mass, illustrated alongside the measured pool deformation during the impact evolution. The inset zooms in onto a timescale of large deformation for the top of the drop, when small differences are most visible. (b) Minimum gas film thickness across the contact region at regular discrete points in time during the bouncing dynamics. The while horizontal dashed lines (found at $t \approx 1$ ms for each case) show the values which are extracted as describing the minimum film thickness reported in each case. }
\label{fig:FigApp}
\end{figure}

In order to ascertain the appropriate resolution level for the results in this work, we have focused on balancing accuracy and efficiency, considering several target quantities that encapsulate a full range of detail from highly temporally and spatially localised features such as the gas film thickness to larger scale metrics such as contact time and coefficients of restitution, thereby providing a stringent test for our platform. We summarize our findings in Figure~\ref{fig:FigApp} and Table~\ref{tab:tabApp}, respectively. 

The typical vertical dynamics during impact is shown in Figure~\ref{fig:FigApp}(a), with the top of the drop, its center of mass, and its bottom coordinates being overlaid with the height of the liquid pool just underneath the drop. While our lowest resolution level $8$ (representing $2^8$ grid cells per dimension for our domain spanning $20$ drop radii in each dimension) has small deviations compared to its more refined counterparts (as highlighted in the magnified inset), we overall find excellent alignment across resolution levels for the droplet dynamics, translating into suitably converged values for traditional bounce metrics such as contact time and coefficients of restitution. The horizontal dynamics ($z$ versus $t$) showcases very similar behaviour upon inspection. This comparison gives us confidence to explore the contact dynamics more carefully towards uncovering explanatory mechanisms and trends that connect to the experimental findings. Figure~\ref{fig:FigApp}(b) displays the minimum thickness obtained across the contact surface, with a sharp decrease as the drop approaches the pool, followed by a plateau-like behaviour as the drop is in its descent phase, during which the reported minimum is extracted (shown as white dashed lines in the respective figure). Comparing the two panels allows us to observe the descent and ascent stages respectively, including occasionally noisy measurements, particularly at lower resolutions. This noise has encouraged us to formulate the post-processing algorithm mentioned in the discussion of Figure~\ref{fig:Fig4}(b), based on ensuring the reported minimum thickness is maintained over sufficiently long spatial and especially temporal scales. Through numerical experimentation it was found that building a criterion based on the average over a rolling window spanning $5$ adjacent contact grid cells leading to a measured film thickness maintained over $10\%$ of the contact time yielded informative results with limited impact from localised numerical artifacts. Across our studies, it is within the descent period (when the drop presses onto the pool surface and the horizontally moving pool also acts to thin out the gas layer) that the overall minimum is found at all resolution levels. 
There is however rich nonlinear dynamics exhibited throughout the contact period as the competing mechanisms shift during impact and rebound.

\addtolength{\tabcolsep}{-1pt}
\begin{table}
\small
\centering
\begin{tabular}{cccccccc}
\hline
\makecell{\textbf{Resolution}\\\textbf{level}} &
\makecell{CPU \\ runtime (s)} &
\makecell{Grid cell\\count} &
\makecell{Min. grid \\ cell size ($\mu$m)} &
\makecell{Min. gas film\\ thickness ($\mu$m)} &
\makecell{Vertical \\CoR ($\alpha$)} &
\makecell{Horizontal \\CoR ($\epsilon$)} &
\makecell{Contact time\\$t_c/t_{\sigma}$} \\
\hline\hline
8 & $1.90 \cdot 10^4$ & $1.0-1.5 \cdot 10^5$ & $17.969$ & $1.127$ & $0.254$ & $0.517$ & $4.848$ \\
\hline
9 & $1.85 \cdot 10^5$ & $4.2-5.6 \cdot 10^5$ & $8.984$ & $0.473$ & $0.251$ & $0.538$ & $4.919$ \\
\hline
10 & $3.46 \cdot 10^6$ & $1.8-1.9 \cdot 10^6$ & $4.492$ & $0.231$ & $0.252$ & $0.541$ & $4.917$ \\
\hline
11 & $3.12 \cdot 10^7$ & $8.0-8.5 \cdot 10^6$ & $2.246$ & $0.188$ & $0.253$ & $0.542$ & $4.915$ \\
\hline
\end{tabular}
\caption{Direct numerical simulation resolution study for the impact and subsequent bounce of a $2$ cSt silicone oil droplet with $R=0.23$ mm ($Bo=0.024$, $Oh=0.028$), $V=60.3$ cm/s, and $U=15$ cm/s, as the prototypical test case in our study. Here we highlight performance metrics (CPU runtime, grid cell count), geometrical information (minimum grid cell size) and the convergence properties of metrics of interest such as the horizontal ($\epsilon$) and vertical ($\alpha$) coefficients of restitution, the normalised contact time, and the minimum gas film thickness.}
\label{tab:tabApp}
\end{table}
\addtolength{\tabcolsep}{1pt}

In light of the above, we conclude that an $O(\mu m)$-sized minimum grid cell setup enables us to reconcile our approach with our objectives, with resolution level $10$ deemed to be a suitable compromise between robustness and computational cost across our numerical campaign. While careful to differentiate between quantitative predictive capabilities (e.g. for the bounce metrics) and qualitative insight (e.g. for the gas film thickness behavior), we propagate this setup to our parameter studies in Section~\ref{sec:Results}, while also setting it as a template for the GitHub code repository associated with our investigation.

\begin{appen}
\noindent {\bf Supplementary Material.} Supplementary movies are available in \href{https://github.com/rcsc-group/BouncingDropletsMovingPool3D/tree/main/SupplementaryMovies}{the associated repository}.

\noindent {\bf Funding.} The authors gratefully acknowledge the financial support of the National Science Foundation (NSF CBET-2123371) and the Engineering and Physical Sciences Research Council (EPSRC EP/W016036/1).

\noindent {\bf Declaration of Interests.}
The authors report no conflict of interest.

\noindent {\bf Data availability.}
The DNS implementation, as well as associated data pre- and post-processing scripts are made available at \url{https://github.com/rcsc-group/BouncingDropletsMovingPool3D}. Data is available from the corresponding author upon request.

\end{appen}

\bibliographystyle{jfm}
\bibliography{jfm}

\end{document}